# Understanding Growth Mindset Practices in an Introductory Physical Computing Classroom: High School Students' Engagement with Debugging by Design Activities

Luis Morales-Navarro, University of Pennsylvania, luismn@upenn.edu
Deborah A. Fields, Utah State University, deborah.fields@usu.edu
Yasmin B. Kafai, University of Pennsylvania, kafai@upenn.edu

**Background and Context**: While debugging is recognized as an essential practice, for many students, encountering bugs can generate emotional responses such as fear and anxiety that can lead to disengagement and the avoidance of computer programming. Growth mindsets can support perseverance and learning in these situations, yet few studies have investigated how growth mindsets emerge *in practice* amongst K–12 computing students facing physical computing debugging challenges.
**Objective**: We seek to understand what (if any) growth mindset practices high school students exhibited when creating and exchanging buggy physical computing projects for their peers to solve during a Debugging by Design activity as part of their introductory computing course.
**Method**: We focused on moment-to-moment microgenetic analysis of student interactions in designing and solving bugs for others to examine the practices students exhibited that demonstrated the development of a growth mindset and the contexts in which these practices emerged.
**Findings**: We identified five emergent growth mindset practices: choosing challenges that lead to more learning, persisting after setbacks, giving and valuing praise for effort, approaching learning as constant improvement, and developing comfort with failure. Students most often exhibited these practices in peer-to-peer interactions and while making buggy physical computing projects for their peers to solve.
**Implications**: Our analysis contributes to a more holistic understanding of students' social, emotional, and motivational approaches to debugging physical computing projects through the characterization of growth mindset practices. The presented inventory of growth mindset practices may be helpful to further study growth mindset in action in other computing settings.

## Introduction

Debugging—the process of identifying and fixing errors that prevent an application from working as expected—is an inevitable, intrinsic, and often frustrating part of learning computing. As Papert (1980) remarked, "errors benefit us because they lead us to study what happened, to understand what went wrong, and, through understanding, to fix it" (p. 114). However, debugging is also exacting and can be difficult for students to learn and challenging for teachers to teach (McCauley et al., 2008; Hennessy Elliott, 2023). For many students, encountering bugs can generate fear and anxiety that can lead to disengagement and the avoidance of computing (Scott & Ghinea, 2013; Coto et al., 2022). Yet, few studies have treated debugging holistically, considering the "inextricable relationship between thinking and emotion" in learning to persevere and handle failure (DeLiema et al., 2020, p. 210). Traditional debugging pedagogies involve linear strategies to find bugs in code in constrained programming contexts with timely feedback (Silva, 2011; Luxton-Reilly et al., 2018). These approaches only address isolated cognitive aspects of debugging in limited scenarios, failing to address the "emotional toll" (Kinnunen & Simon, 2010; Coto et al., 2022) that debugging may have on learners in open-ended design contexts.

One way to reframe debugging as an opportunity for learning and to decrease hopelessness and attrition among novices is by promoting a growth mindset (Kinnunen & Simon, 2010). Mindsets organize people's beliefs about themselves and their abilities (Dweck & Yeager, 2019). Students with growth mindsets believe their skills and abilities can evolve and develop over time, whereas students with fixed mindsets believe their intellectual capacity and social characteristics are unchangeable (Dweck, 2006). Mindsets influence emotional states (Heyman & Dweck, 1992; Robins & Pals, 2002), and growth mindset interventions seem to reduce learners' anxiety (Smith & Capuzzi, 2019). Within computing, prior research provides promising evidence of mindsets playing an important role in student learning. This research shows that while students with growth mindsets tend to perceive bugs as opportunities for learning, those with fixed mindsets become frustrated with failure and see it as a challenge to their intelligence (Burnette et al., 2020; Nolan & Bergin, 2016). Thus, promoting a growth mindset in the classroom is important since debugging is such an essential practice, and persevering through debugging is critical to students' learning. Traditional growth mindset interventions in computing education, which tend toward a more didactic lecture/telling approach (Burnette et al., 2020; Simon et al., 2008), have not consistently yielded a significant impact on student performance. Further, actual student practices do not always align with self-reported instrument responses (Stout & Blaney, 2017; Gorson & O'Rourke, 2019). With the additional lack of research attention to K–12 computing students' growth mindset, this calls for research into students' growth mindset *practices*, i.e., students' active behaviors that put growth mindset into practice when responding to bugs in computing.

In this exploratory, observational study, we examine the growth mindset practices that high school students exhibited when they created buggy physical computing projects, in particular electronic textiles (hereafter e-textiles), for their peers to solve and then fixed each other's projects. We positioned learners as bug designers for their peers, potentially increasing their engagement with bugs by allowing them to be creators of bugs rather than just solvers. We report on four cases of student groups participating in our implementation of "Debugging by Design" (DbD), an open-ended design learning activity, in a high school computing classroom over a period of eight 50-minute class periods (Fields et. al., 2021). This activity was situated in the context of Exploring Computer Science, an introductory year-long high school computing curriculum (Margolis et al., 2017). In DbD, students worked with e-textiles, which involve creating wearables using programmable microcontrollers, sensors, and actuators that can be sewn into fabrics (Buechley et al., 2013). Making physical computing projects not only includes designing functional circuits but also writing code that controls interactions. As such, debugging in physical computing involves finding and fixing semantic and syntactic bugs in code as well as design bugs in the circuits of physical artifacts (Searle et al., 2018; DesPortes & DiSalvo, 2019). We observed high school students to see if and when they demonstrated growth mindset practices while designing buggy e-textile projects for their peers and debugging peer-designed projects. Building on traditions of qualitative and learning sciences research in computing education (Tenenberg, 2019; Margulieux et al., 2019), we draw on microgenetic, moment-by-moment, analysis of student team interactions to address the following research questions: (1) What growth mindset practices do students exhibit while designing and solving bugs in physical computing projects? and (2) Where and how do students exhibit growth mindset practices during the DbD activity?

**Background**

*Toward a Holistic Approach to Debugging*

While computing education research has historically focused on cognitive aspects of learning, non-cognitive dimensions such as motivation and emotion are equally instrumental in supporting all students (Lishinski & Yadav, 2019). The latter is the focus of this paper. For instance, while the importance and prevalence of debugging have been recognized in computing education, little attention in research and practice has been given to the role of motivation and emotion in debugging (DeLiema et al., 2020). Indeed, traditional approaches to learning and teaching debugging tend to focus on the conceptual and cognitive challenges that emerge when encountering bugs (Li et al., 2019). Procedure-based linear strategies with flowcharts and tables are some examples (Silva, 2011). Others are self-directed strategies based on domain-specific knowledge or systematic debugging strategies (Michaeli & Romeike, 2019; Jonassen & Hung, 2006). However, debugging is a complex process where troubleshooting, identity, and emotion intersect (Dahn & DeLiema, 2020), generating deep emotional investment (Turkle & Papert, 1990; Coto et al., 2022). Learners' emotional states dominate their perception of their programming experiences (Kinnunen & Simon, 2010), which may affect their performance and motivation (Heyman & Dweck, 1992; Hecht, 2021). As a result, we must consider debugging a holistic experience, as it is shaped by and shapes cognition, emotion, and motivation (DeLiema et al., 2020; Dahn et al., 2020).

Attending the role of motivation and emotion in debugging has historic grounding in computing learning research. For example, in their studies of novice programmers, Perkins et al. (1986) identified a dichotomy of reactions among beginners when they encounter bugs. Some novices see bugs as inherent and exciting challenges of learning to code, while others become frustrated, perceiving bugs as negative reflections of their performance. It is not uncommon for this second group of novice programmers to experience discomfort and anxiety when debugging and encountering failure (Lovell, 2014; Kench et al., 2016; Nolan & Bergin, 2016). These students see bugs as threats to their self-esteem and reputation and subsequently doubt their ability to learn. This perception of debugging experiences as reflecting innate programming ability is related to learner mindsets.

In psychological motivation research, Dweck (2006) has organized opposing self-beliefs about innate and malleable intelligence into "fixed" and "growth" mindsets. Students with fixed mindsets perceive their intellectual abilities and social traits as unchangeable—parallel to Perkins et al.'s (1986) learners who saw bugs as a negative reflection on their capabilities. On the other hand, learners with growth mindsets believe their competencies and abilities can change and be developed over time—comparable with learners who see bugs as inherent challenges and opportunities to learn (*Ibid.*). Learners with growth mindsets tend to have higher resilience in challenging tasks, such as debugging, than those with fixed mindsets (Apiola & Sutinen, 2020; Hetch, 2021; Yeager & Dweck, 2012). Indeed, mindsets influence how people choose challenges, persist over setbacks, and create value judgements (Dweck & Yeager, 2019). Mindsets are malleable and can be changed through small interventions in which students learn about the potential to change their abilities and how to develop growth mindsets (Apiola & Sutinen, 2020; Yeager et al., 2012). However, the effect of an intervention may depend on the psychological affordances of the context, that is, the learning environment characteristics that support growth mindset beliefs (Hecht et al., 2021). This opens opportunities to study how growth mindsets can be promoted within computing learning environments.

### *Growth Mindsets in Computing Education*

In computing education, growth mindsets have gained relevance in research on motivational and emotional factors contributing to debugging (Kwak et al., 2022). For instance, Murphy and Thomas (2008) argued that in computing, while students with growth mindsets perceive bugs as opportunities for learning, those with fixed mindsets become frustrated with failure and see it as a challenge to their intelligence. When encountering bugs and attempting to fix them, novice learners may experience feelings of helplessness, frustration, and anxiety, promoting fixed mindsets (Scott & Ghinea, 2013; Nolan & Bergin, 2016), leading to withdrawal and even abandonment of computing (Margolis et al., 2017). Yet while fixed mindsets are prevalent and common among novice computing learners (Gorson & O'Rourke, 2019), designing learning environments and interventions that promote growth mindsets can positively influence student motivation and engagement (Flanigan et al., 2022).

The existing, but limited, research on growth mindset in computing education has yielded some promising results that suggest that mindset interventions increase interest, particularly among novices (Burnette et al., 2019). However, most studies have been conducted at an undergraduate level (Flannigan et al., 2022; Kwak et al., 2022; Apiola & Sutinen, 2020; Burnette et al., 2020; Quille & Bergin, 2020; Woods, 2020; Gorson & O'Rourke, 2019; Stout & Blaney, 2017; Nolan & Bergin, 2016; Scott & Ghinea, 2013) or with adults (Rangel et al., 2020), with little attention given to younger middle and high school students (Loksa et al., 2016; Kench et al., 2016; Margolis et al., 2017). At the same time, the majority of these interventions use didactic techniques, such as delivering growth mindset information through videos, teacher lectures, and readings, occasionally followed by reflective writing and "saying is believing" exercises (Simon et al., 2008; Cutts et al., 2010; Rangel et al., 2020). Most of these efforts prioritize top-down approaches to inform students about growth mindset rather than facilitating learning experiences that support students' active development of growth mindset in practice. Some exceptions include interventions that use problem-solving checklists for students to track their programming progress in projects (Loksa et al., 2016) and goal-setting interventions where students plan, follow through, and reflect on their own goals (Woods, 2020).

Yet beyond information delivery or even problem-solving checklists or goal setting, it is critical to foster growth mindsets in actual classroom practice (Dweck & Yeager, 2019). For instance, in mathematics, Campbell et al. (2020) identified *practices* that characterize growth mindsets in learning activities and situated these within larger learning theories. They distinguished six categories of practices in which learners with growth and fixed mindsets behaved in opposing ways: challenges, persistence, effort, praise, learning goals, and the success of others. For example, in dealing with challenges, learners with growth mindsets choose challenges that allow them to learn more, while those with fixed mindsets avoid challenges in areas that they cannot ensure they will "do well" (Campbell et al., 2020). Similarly, Haimovitz and Dweck (2017) examined how adult and socialization practices foster both growth and fixed mindsets in children, highlighting practices for responding to success and failure in process-focused teaching. Centering explicitly on *practices* may be essential because, with traditional growth mindset interventions in computing education practices do not always align with self-reported instrument responses (Stout & Blaney, 2017; Gorson & O'Rourke, 2019). Research is needed on designing computing learning environments that promote growth mindset *practices*, resilience, and persistence in debugging. This kind of research on observable growth mindset practices in computing education may be helpful for teachers to identify and promote student behaviors that may lead to a growth mindset.

In sum, alongside conceptual understanding and debugging practices, students need to engage with growth mindset practices and develop growth mindsets to overcome the motivational and emotional challenges they encounter when facing bugs. While top-down approaches and exposure to models of growth mindset can be helpful, experience-based mindset interventions may be better suited to supporting students to overcome challenges in computing (Burnette et al., 2020; Margolis et al., 2017). In the remainder of this paper, we focus on an intervention that was designed to engage students in creating intentional bugs for their peers to solve and promote their growth mindset practices.

### *Growth Mindset Practices in Debugging by Design*

Building on this research base, we investigate growth mindset in action by analyzing high-school-aged students' growth mindset practices when engaging in DbD, an intervention where students design and then exchange buggy e-textiles projects to solve. Design-based physical computing learning activities, such as e-textiles, may support the development of growth mindsets because they emphasize the *process* of learning over its outcomes and the importance of personal *effort* in connection to personal interests (Vongkulluksn et al., 2021). Physical computing activities that focus on fostering creativity and designing interest-driven artifacts are particularly effective in promoting student motivation (Przybylla & Romeike, 2018). DbD is a good case study for studying growth mindset practices in computing because it is a student-centered intervention designed to empower students in debugging by designing creative, multimodal buggy projects for others to solve. In this case, we anticipated that students would feel greater control over bugs because they chose and created them rather than running into them accidentally, as normally happens in designing computational artifacts.

E-textiles learning activities provide a rich context to design and solve bugs, especially ones that integrate code errors with circuitry problems, forcing learners to trace, evaluate, and isolate problems across modalities (Hodges et al., 2020). As such, debugging physical computing artifacts involves addressing syntactic, semantic, conceptual, and design issues as bugs may emerge in code and circuit/artifact design (DesPortes & DiSalvo, 2019; Jayathirtha et al., 2018; Lui et al., in press; Schneider, 2022). For example, semantic errors may include variables not being declared, type mismatches, or method calls with wrong arguments; syntactic errors may include missing delimiters; conceptual bugs may include intentionality bugs that occur when learners expect programs to behave in ways beyond the code given; and circuitry design bugs may include short circuits and wrong connections (Jayathirtha et al., 2018; DesPortes & DiSalvo, 2019; Fields et al., 2021; Booth et al., 2016). Students use a variety of debugging strategies when they encounter bugs in physical computing systems, such as isolating problems through forward and backward reasoning, formulating hypotheses, generating solutions, and testing hypotheses and solutions (Jayathirtha et al., 2020). The hybrid nature of physical computing allows for rich debugging learning as students develop troubleshooting and pragmatic skills (Jayathirtha et al., in press; Lui et al., in press).

Contrary to traditional growth mindset interventions (e.g., "saying is believing" exercises), which often fail among adolescents because they do not align with students' enhanced desire to be "treated as though they are competent, have agency and autonomy, and are of potential value to the group" (Yeager et al., 2018, p. 104), in DbD learners are in control of designing bugs in interest-driven, personalized projects. In DbD, students are in charge of intentionally creating the bugs rather than just stumbling upon them. As such, they decide what kind of bugs to select and make, drawing on their own prior experiences as well as those of their classmates. DbD was designed to promote growth mindsets by putting learners, rather than teachers or researchers, in charge of creating intentional problems, turning bugs into a feature of the learning product rather than a stumbling block. An early study based on students' self-reports suggested that when students consciously created bugs for peers, they improved their abilities to detect and fix bugs, collaboratively solved problems, and increased their confidence in debugging. Although some students expressed frustration when solving bugs created by their peers, afterwards they reported feeling more comfortable and competent in solving and designing bugs (Fields et al., 2021). Our research on DbD in this exploratory, observational study focuses on the growth mindset practices students exhibited when designing bugs for others and solving bugs, as well as when and where these practices emerged in context.

**Methods**

*Site and Participants*

We conducted this study in Spring 2019 at a high school introductory computing class in a metropolitan school district on the West Coast of the United States. In the class, 25 students ages 14–18 received parental consent and assented to participate in the study: 11 self-identified as female, and 14 self-identified as male. In 72% of cases, participating students spoke a language other than English at home; 80% had no prior computing education experience (no CS classes or workshops before this course); and 80% had family members with at least some college education. The class was ethnically and racially diverse, with 48% of participating students identifying as Latinx, 36% as Asian American/Pacific Islanders, 8% White, 4% other, and 4% not reporting their race or ethnicity. Ben, the teacher, had three years of experience teaching e-textiles and collaborated with the research team to develop the curriculum.

We requested Ben's help in choosing four groups as case studies for close observation, asking for a range of student groups so that we could have a diverse view of how students responded to the DbD activity. Drawing on his knowledge of student groups who had just completed the third (collaborative) project in the e-textiles curricular unit, Ben selected four student groups (n = 10 students) for case studies to represent a range of student interaction and performance styles. Two groups were very high performers: students finished projects on time and collaborated smoothly as a team. The two other groups struggled in prior projects: students did not complete projects on time, had trouble getting projects to work, and had more awkward collaborations (e.g., major disagreements and different working styles). Case study groups included two groups of two students (Evelyn and Nicolás; Liam and Sophia) and two groups with three students (Lucas, Emma, and Lily; Georgia, Gabriel, and Camila) (see Table 1 for demographic characteristics of participants).

Table 1. Self-reported demographic characteristics of participants. All names are pseudonyms.

| Group | Participant | Gender | Race/Ethnicity | Previous Computing Education |
|---|---|---|---|---|
| 1 | Lucas | Male | Asian | No |
|  | Emma | Female | Latina | No |
|  | Lily | Female | Latina | No |
| 2 | Georgia | Female | Asian | No |
|  | Gabriel | Male | Latino | No |
|  | Camila | Female | Asian | Yes |
| 3 | Evelyn | Female | Latina | No |
|  | Nicolás | Male | Latino | No |
| 4 | Liam | Male | Asian | No |
|  | Sophia | Female | Asian | No |

*Debugging by Design*

We situated DbD within the e-textiles unit of Exploring Computer Science (ECS). This is a year-long inquiry-based computing curriculum consisting of multiple units committed to broadening participation in computing through a building talent approach that addresses the structural inequities and belief systems that limit participation from historically marginalized groups (Margolis et al., 2017). Within ECS, the promotion of growth mindsets among students and teachers already plays an important role, as it puts equity first by considering that all students with access to quality education can grow in engagement and capacity (Margolis et al., 2017). However, the e-textiles unit is the only unit in the ECS curriculum that prioritizes having students design personally relevant projects (Fields et al., 2018; Fields & Kafai, 2023).

The unit[1] centers around the design of the four e-textiles projects—paper card (simple circuit), wristband (parallel circuit with switch), interactive mural (programmable circuit with two switches), and human sensor (computational circuit with analog sensor)—that support students in learning computing, electronics, and crafting while designing personally relevant artifacts (for design principles of the unit, see Kafai & Fields, 2018; for evaluations of student learning, see Kafai et al., 2019; Fields et al., 2021). The unit requires students to apply concepts such as sequences, loops, conditionals, and variables (learned in Unit 4) to a text-based programming language (Arduino). In addition, students learn new programming concepts such as nested conditionals, data input from sensors, and functions.

DbD was intentionally situated between the third and fourth projects in the e-textiles unit, after the collaborative class mural described above. This allowed students to apply their experiences with bugs from earlier projects to their DbD designs and apply any knowledge gained from DbD to their final project. The activity took place during eight 50-minute class sessions over a period of two weeks (see Table 2). During the first session's "Hall of Problems," students first discussed with their partners, then the whole class, different errors and problems they encountered when working on e-textiles projects, creating a public, collective set of all the bugs they could think of that occurred in their prior designs. Following, during session two, student groups created a select list of bugs to include in their DebugIt along with an aesthetic design or sketch of the project. The teacher encouraged students to be considerate of their peers in putting only as many bugs in as could be solved in a single class period, which led to some careful selection from several initial bug ideas. In sessions four and five, and after receiving approval of their design from their teacher, students created their buggy designs, or DebugIts, which required them to craft a non-functional project with its circuit and code (see Figures 1, 3, and 4). In the following two sessions, students exchanged buggy projects with their peers and solved each other's projects. During the last session, pairs presented their solutions (sometimes full, sometimes incomplete) to the class and reflected on the strategies they used to solve the DebugIts.

Table 2. Debugging by Design Activity.

---

[1] The entire curriculum can be found at http://exploringcs.org/e-textiles

| | |
|---|---|
| **Class 1** | **"Hall of Problems":** As partners, then as a whole class, students list e-textile problems. Then they categorized these problems into groups, which are written on posters on the classroom walls. |
| **Class 2** | **DebugIt Design:** Students plan their DebugIts, turning in a list of problems with solutions as well as a circuit diagram showing any circuitry bugs. The teacher had to approve designs. Most groups revised their designs after teacher feedback, which continued into Class 3. |
| **Classes 3-5** | **DebugIt Construction:** After receiving teacher approval on their design, students created their DebugIts: sewing and coding their projects with intentional bugs. They also wrote an intention statement for how the project ought to work. |
| **Classes 6-7** | **DebugIt Solving: Under the teacher's direction, students exchanged projects and had 1.5 class periods to solve them as best they could.** Students individually, then in small groups, reflected on what the best, most frustrating, and surprising parts of the entire debugging by design experience were. |
| **Class 8** | **Reflection on Problem-Solving Strategies:** Individually, then as pairs, and finally as a class, students reflected on the kinds of strategies they used to solve DebugIts. They shared strategies at the front of class, such that DebugIt's designers could respond to any bugs the solvers missed or note any bugs unintentionally put into the design. |

*Data Collection*

To investigate the growth mindset practices that students exhibited and the situations, challenges, or contexts where these practices seemed to emerge, we drew on: 1) in-class observational field notes; 2) videologs (based on video recordings) of participants working on their projects; 3) artifact collection; and 4) interviews with students (n = 10). Throughout the eight class sessions of DbD, we positioned video cameras in the four selected student groups. In addition, one researcher rotated across the groups and took observational field notes and regular photographs of student work, occasionally asking the groups about what they were doing and why. We logged and annotated the video recordings by adding information regarding what students did in the video, their design and debugging processes, and their emotional reactions. In addition, we collected all student products: circuit diagrams, code, and pictures of the artifacts designed by the students. This multi-source approach provided a more complete picture of what happened in the classroom interactions, which was supplemented by post-interviews with student groups several weeks after the DbD activity ended. All the data was gathered and organized by group to put together a picture of how each group engaged with the activity.

*Data Analysis*

To answer the research questions, (1) What growth mindset practices do students exhibit while designing and solving bugs in physical computing projects? and (2) Where and how do students exhibit growth mindset practices during the DbD activity?, we focused on moment-to-moment microgenetic observational analysis of student team interactions in designing and solving bugs for others. By examining the process of learning and the experiences of learners, we built on a longstanding tradition of the learning sciences in computing education research (Margulieux et al., 2019). Moment-to-moment analysis, also known as microgenetic analysis, is a common method in the learning sciences that allows researchers to "observe learning processes as they occur and do so in such a way as to permit strong inferences about learning processes" (Chinn & Sherin, 2014). This type of analysis is well suited to investigating learning in small collaborative groups (Taylor & Cox, 1997) and especially the moment-to-moment co-occurrence of processes in a learning environment, such as development practices while completing a learning activity (Schoenfeld et al., 1993; Chinn & Sherin, 2014). In computing education research, moment-to-moment analysis can be especially useful for studying learners' non-cognitive and cognitive practices. Because microgenetic research necessitates methods that attend to the rich processes that take place while students learn, qualitative research and case studies are frequently used to distill the practices and processes that students engage with and to explain how learners engage with them in context (Chinn & Sherin, 2014).

In this observational study, we used qualitative methods to identify practices in action through a ground-up analysis that enabled us to build on existing theory while being open to practices that may not have been observed in other contexts. By practices, we mean observable behaviors, mostly discernible during conversation or in actions aimed at fixing bugs, such as editing code or making visible changes to a project or design. Building on traditions of qualitative methods in computing education research (Tennenberg, 2019), we analyzed the data in three steps. First, we read data sources from two student groups chronologically focused on inductive descriptive coding (Saldaña, 2015). Through this process, we developed a collection of codes for instances, moments, or excerpts where students exhibited growth mindsets in observable ways. Salient codes included moments of praise, peer support, sharing failure, valuing the success of others, setting learning goals, not being afraid of being wrong, and moments of challenge. This analysis helped us understand and identify how students practiced a growth mindset during the learning activities. We then clustered the descriptive codes into similar categories to detect patterns and interrelationships. Previous studies on mindset practices and behaviors (Campbell et al., 2020; Dweck, 2006; Dweck, 2000) informed the clustering process. Taking an inductive approach first to see what emerges from the data and then clustering codes according to the literature is a common practice in qualitative research in computing education (Tennenberg, 2019; Searle & Kafai, 2015). In our case, this was important since the literature on growth mindset practices has largely been built in other fields (i.e., social psychology and mathematics education), and such inductive coding served to openly investigate how growth mindset practices emerged in a high school physical computing context prior to building a coding scheme informed by the literature. After developing the coding scheme (all coding categories are explained with examples in the findings section; for a table with the codebook, see Appendix 1), we discussed it with two researchers familiar with the data, received feedback, and revised it. In a second round of analysis, we applied the coding scheme across all groups through a deductive reading (Saldaña, 2015).

Following that, in order to gain a comprehensive understanding of the contexts in which students engaged in growth mindset practices, we created narrative-based accounts that situated the practices previously identified in the moment-to-moment richness of the learning context (Schoenfeld et al., 1993). The research team drew on the chronological assemblage of data for each group to create narratives that included contextual details that influenced growth mindset practices, such as peer interactions (within and between groups), teacher support, and class discussions. This enabled us to further analyze the narrative accounts to uncover when and how growth mindset practices emerged during the DbD activity.

**Findings**

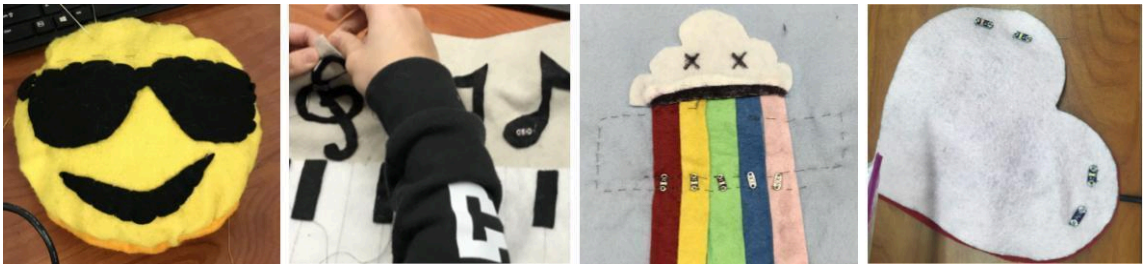

*Figure 1*. Projects created by the four focal groups. From left to right: an emoji plushie created by Sophia and Liam; a piano created by Georgia, Gabriel, and Camila; a sick cloud throwing up a rainbow created by Evelyn and Nicolás; and a heart created by Lucas, Emma, and Lily.

Our analysis identified five growth mindset practices: (1) choosing challenges that lead to more learning; (2) persisting after setbacks; (3) giving and valuing praise for effort; (4) approaching learning as constant improvement; and (5) developing comfort with failure. The first four practices coincide with the practices proposed by Campbell et al. (2020); the last practice emerged inductively from analysis. In the following sections, we introduce these five practices and highlight where they emerged during the various aspects of the activity (planning, designing, and debugging) and across individual or peer-to-peer encounters. Finally, we share two in-depth case studies demonstrating how these practices emerged in classroom and small-group contexts.

***Growth mindset practices***

*Choosing challenges that lead to more learning*

Throughout DbD, students purposefully chose challenges that led them to learn new things, "throwing themselves wholeheartedly into difficult tasks—and sticking with them" (Dweck, 2000, p. 3). As an example, when planning their DebugIt, Evelyn explained she wanted it to include music that could be activated when people pressed a button. Nicolás replied, "That's too hard" because they had no previous experience coding music. Yet, he turned to his computer, and after a few minutes of googling and tinkering with Arduino sample code, he showed Evelyn how they could make the Arduino board play notes, learning something new in the process (Lesson 2, Video). In a different group, when making a DebugIt, Liam suggested to his partner Sophia that they make an emoji plushie, a three-dimensional e-textiles project with the electronic components distributed inside the artifact and on its surface (Lesson 2, Video) (see

Figure 1). While they had never made a plushie, they decided to place lights on the front and the microcontroller on the back. Thus, knowing they had no experience in making three-dimensional e-textiles, the group took on the new challenge to arrange the electronic components in this spatially distributed way and, mischievously, to embed a bug inside the plushie. These are just two of the 45 instances we identified in which students consciously chose challenges that led them to deeper learning.

*Persisting after setbacks*

Setbacks are common in learning computing and engineering and are an essential part of debugging that can cause frustration and, at the same time, provide opportunities for students to persist. Frustration as part of the learning process is meaningful because it creates opportunities for students to persist and, as Dweck (2006) writes, "to turn an important setback into an important win" (p. 93). Indeed, although frustration can be unpleasant (Baker et al., 2010), it can precede successful learning (D'Mello, 2013). For this practice, we identified two subcategories: (1) how frustration sometimes precedes persistence, and (2) how frustration co-occurs in instances of persistence.

     As an example of frustration preceding persistence, we share a time when Evelyn was frustrated with fixing a circuitry bug in her sewing. She was overwhelmed, exclaiming, "This is too much... I don't know what's wrong with it." She was upset, feeling like the problem was "too much.". Her partner, Nicolás, encouraged her, but she insisted she didn't know how to fix it. Nicolás offered to help: "déjame [let me] look at it" (Lesson 7, Video). Twenty minutes later, and after working on other aspects of the project, Evelyn persisted and returned to the bug, folding and bending the artifact to get it to work until the lights turned on. During this process, Evelyn discovered that the problem was in the loose connections of the conductive thread. "I could have fixed it a long time ago if I knew the connections were all loose," she reflected (Lesson 7, Video).

     On a different occasion, Evelyn seemed upset dealing with a bug, and Nicolás encouraged her: "You can fix it! Fix it!" (Lesson 7, Video). In this instance, frustration and persistence co-occurred. Despite her frustration, Evelyn managed to get the lights to turn on. Thus, with encouragement from her partner, as Evelyn persisted, she evaluated and analyzed problems, looking for alternative ways to solve them. These two examples illustrate the occurrence and co-occurrence of frustration and persistence and how setbacks, although frustrating, became opportunities for learning in the process of creating and solving e-textile bugs.

*Giving and valuing praise for effort*

While working on their DebugIts, students praised each other's efforts, strategies, and processes. This kind of praise may foster mastery-oriented responses when facing difficult problems (Dweck, 2000). When learners receive supportive responses from peers, they are more likely to accomplish their tasks and participate in generating new ideas (Jordan & McDaniel, 2014). As an illustration, while Georgia struggled with her sewing, Camila encouraged her, "You're almost there; keep going," and when they finished stitching, a connection celebrated their effort, saying, "We got it!" (Lesson 3, Video). Nicolás even praised Lily for adding mischievous and nuisance code, such as declaring four extra variables that were not used in the project just to confuse the debugging team, saying, "Good job, you made it really hard [for the other team]" (Lesson 7, Video). These two instances showcase the frequent praise and encouragement for effort that students gave each other in designing and solving bugs. This kind of support from peers, centered around effort, promotes a growth mindset by framing setbacks and challenges as inherent aspects of learning that can be overcome with persistence.

*Approaching learning as constant improvement*

Student teams also embraced learning as constant improvement and looked for ways to improve their projects and their competence. As Dweck (2000) argues, this reflects a desire to learn, get smarter, and acquire new skills. We see this when, after getting all the lights of the project to work, Liam decided to "resew the thing" to make sure that all the connections were tighter and better organized because "the threading is kind of bad" (Lesson 6, Video). In another episode, after working on the code and changing the pins used, Sophia decided to redraw the circuit diagram. She could just have annotated or fixed the old one but instead wanted to make a new one "to match the code" (Lesson 6, Video). In these two examples, we can see how they sought opportunities to improve their performance and saw the goal of learning as constant improvement and not just showcasing or achieving a certain level of performance.

*Developing comfort with failure*

Developing comfort with failure generates opportunities for learners to embrace and overcome mistakes with proper motivation and guidance (Dweck, 2000). When learners are worried about failure, they are more likely to develop a fixed mindset and question their ability to learn. On the other hand, learners with a growth mindset "see their own failures as problems to be solved, and they see other people's failings that way as well" (Dweck, 2000, p. 88). In DbD, we saw evidence of students' comfort with failure when: (1) asking questions and requesting feedback without fear of being wrong; (2) sharing failure with their peers; and (3) embracing failure and imperfection as part of the process. Sophia, for example, asked questions to her peers without being worried about being wrong. While sewing a circuit, she asked Liam, "Can positive and positive cross each other?" (Lesson 3, Video). Later, Liam asked Sophia if the constant pin values in Arduino ("LOW" and "HIGH") could be lowercase (Lesson 3, Video). These are two of many instances in which students, including high-achieving students like Liam and Sophia, requested feedback from peers without any fear of being wrong. Peer support when facing uncertainty can generate a safe environment where students are not afraid of asking for help from others (Jordan & McDaniel, 2014).

Further, when learners encountered failure while working on their projects, it was not uncommon for them to openly share failure with their peers. As an example, Evelyn shared a mistake she made when crafting the circuit: she said, "Oh my god! I was going to keep going without sewing up the light!" when she noticed a gap in sewing and laughed at her own mistake (Lesson 4, Video). On a different occasion, she showed Nicolás how the thread got stuck because she "messed up with the glue earlier" (Lesson 4, Video). Similarly, Lucas shared a mistake he made when writing conditionals (Lesson 3, Video; more details in the case study). These are instances of seeing failure as something worth sharing.

Furthermore, students embraced imperfection and failure as a feature of their projects—separate from the intentionally designed bugs. For instance, when Emma noted that the two heart-shaped pieces of felt from the group's project were not perfectly aligned, she described them as "a little bit off, but it's okay.". She also went on to offer a simple, imperfect solution, saying, "We'll just cut it off; it doesn't have to be perfect" (Lesson 4, Video). These are some examples of how students sought help without fear of being wrong, "messing up," making mistakes, or having imperfect projects. Developing comfort with failure helped students reframe failure as a setback they could overcome and as an opportunity for learning.

*Growing mindsets across groups*

The four groups we analyzed engaged in the five identified growth mindset practices in 231 instances. Figure 2 shows the frequency of instances of each practice by group of interest (Chart A) as well as information about the context of each instance, when they occurred (e.g., when planning the project; Chart B), and who was involved (e.g., teacher and learner; Chart C). Students engaged with growth mindset practices through peer-to-peer and learner-teacher interactions throughout the DbD activity, including when planning their projects, making buggy projects, and debugging their peers' projects. More than half of the instances of growth mindset practices identified (56%) occurred when learners made buggy projects for their peers to solve (this is not surprising since students spent three 50-minute sessions designing DebugIts). Most of the instances coded (78%) occurred in peer-to-peer interactions, although we also identified growth mindset practices when students worked alone or in learner-teacher interactions. The frequent occurrence of growth mindset practices while designing buggy projects and primarily during peer-to-peer interactions is showcased in the case studies below.

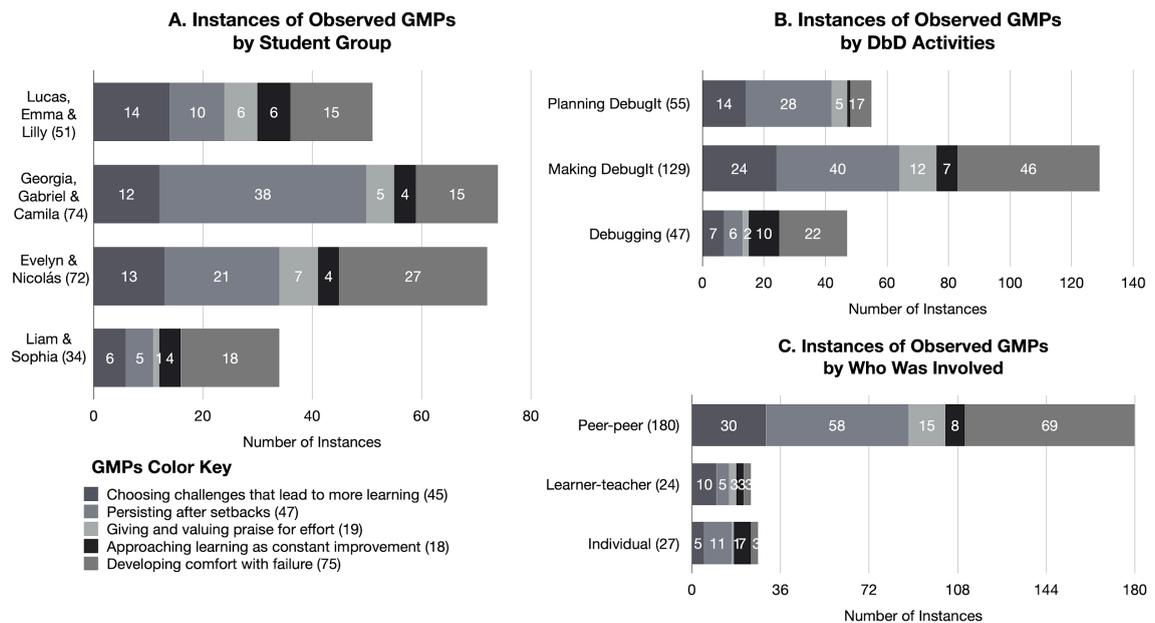

*Figure 2.* Charts of instances of growth mindset practices (GMPs) by student group (Chart A); by DbD activities (Chart B); and by who was involved (Chart C). The key indicates the practices observed.

### *Case Study 1: Georgia, Gabriel, Camila, and the Musical Keyboard*

One case study group—Georgia, Gabriel, and Camila—illustrates how the five growth mindset practices described above came up in the context of the DbD activity. Gabriel, who, according to the teacher, tended to be off task in the previous term, was purposefully grouped with high-achieving students. Together, they created a felt piano that played music and lit up, which included five circuitry/crafting bugs (adding unnecessary thread, reverse polarity in a circuit connection, and a bad/loose connection) and four coding bugs (that included semantic issues: passing the wrong parameters to a function, messing up the sequence of the desired light pattern, and one syntactic problem: missing a delimiter).

We identified 75 instances in which the group showcased growth mindset practices (62 during peer-to-peer interactions and 12 during learner-teacher interactions), even from the initial planning stages. As the group brainstormed what they wanted to make for their DebugIt and the bugs they wanted to include (see Figure 3), Camila suggested an area that went beyond their current expertise: making a piano and coding music. Gabriel hesitated, "We've never used music before, and we have two days" (Lesson 2, Video): they had not learned how to code music in class and did not have much time to figure it out by themselves, but they did have access to a supplementary instructional guide on e-textiles with music. Ben, their teacher, who overheard the conversation about coding music, encouraged the team to do it and even to create a bug related to music. They decided to play *Twinkle, Twinkle, Little Star* and to have Gabriel code it. This required understanding how to play notes by using arrays to define pitch and duration in numerical values and using new functions (tone() and notone()) in the Arduino library in combination with for loops to iterate through the arrays and play the music. Although Gabriel shared that he had "no clue how to code music," he took charge of this new challenge and learned about frequencies and tempo with an instructional guide. Thus, the group admitted their lack of expertise and chose a challenge that would lead to new learning. Of note, it was in the context of both peer and teacher support that Gabriel decided to take on this new challenge.

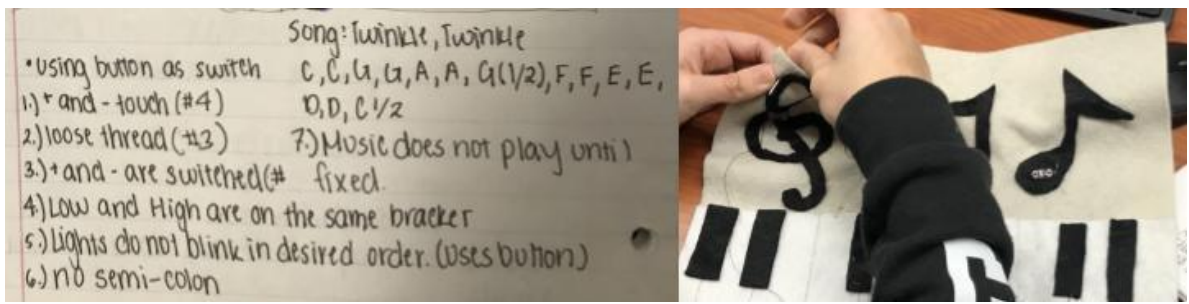

*Figure 3.* Georgia, Gabriel, and Camila's DebugIt: (1) Left: list of bugs included in their project, plus order of tones; (2) Right: Georgia sews musical notes above the keyboard.

Camila and Georgia worked together on crafting the keyboard, and Georgia became frustrated when gluing the musical notes: "I can't do this; I physically can't do this" (Lesson 3, Video). It was hard for her to calculate how much glue to apply to the felt. Having poured glue all over the project, she complained, "I don't like glue; now it's a mess." Georgia's frustration created an opportunity for her to persist and overcome this setback with the support of her peers. Camila provided just-in-time praise for effort to motivate her to keep going: "You're almost there, Georgia; that looks very good" (Video, Lesson 3). In persisting, Georgia tried different strategies: first using a piece of felt to apply glue and then using the tip of the bottle to spread it. The latter strategy worked, and satisfied, she praised her own work: "This [keyboard] looks so good!" (See Figure 3). Here, two growth mindset practices worked together: frustration generated opportunities for persistence, and just-in-time praise for effort helped reframe setbacks as inherent aspects of learning that can be overcome with persistence.

As work on the project continued, Georgia, who had little experience sewing and crafting, shared her mistakes when taking on a challenge that led to more learning. "I'm just warning you, I'm not the best at sewing," she said as she started sewing the circuit. Not afraid of failure, Georgia shared her problems with Camila, who helped her tie knots and make her connections tight. They shared their failures with each other and laughed together about the unintentional mistakes they made while designing their buggy project. Camila, for example, shared that she sewed an LED to the wrong pin of the microcontroller, and Georgia shared that she sewed an LED backwards. Comfortable with failure, when students chose challenges that led to more learning, they embraced mistakes and overcame them.

When coding *Twinkle, Twinkle, Little Star*, Gabriel persisted after encountering many setbacks and frustrations. He did not know how to trigger the song to play, but after fixing many unintentional semantic bugs in his code (for instance, iterating over the pitches array and not the duration array) and getting support from the teacher, he managed to make the microcontroller play music. When it finally worked (albeit not the way he wanted), with frustration, he said, "I have no clue why it sounds so hideous; this is devilish trap music" (Video Lesson 4). His peers did not like the sound. Yet he persisted and continued working on improving the song by fixing the pitch values in the array. The music became an Easter egg in the project; they used the light sensor on the microcontroller to trigger the song. This idea brought in new challenges and unintentional bugs because it was the first time they used a light sensor. With the help of a tutorial, Gabriel managed to read sensor data but struggled to understand that sensor readings went from 0 to 1024 and were not just "LOW" and "HIGH" like when working with switches. After struggling with the conditional statement for the sensor, with the help of a researcher, he got it to work. Gabriel explained: "If you take the flash on your phone and put it on your CP [microcontroller], it'll make sounds happen." This is an example of how when students chose challenges that led to more learning, their frustration became an opportunity to overcome setbacks, persisted when things did not go as expected, and approached learning as constant improvement. When Gabriel was done with the code, he insisted they had to test it to see if it worked and, if not, fix it. They tested several times, and every time, together, the group came up with new ideas of how to improve the project.

*Case Study 2: Lucas, Emma, Lily, and the Buggy Heart*

In another case study group, Lucas, Emma, and Lily, a group chosen by the teacher for their tendency towards lower performance earlier in the class, engaged with growth mindset practices in 56 observed instances (36 during peer-to-peer interactions and 5 during learner-teacher interactions). Together, they created a heart with two sheets of felt, one over the other, playfully hiding some bugs in between the pieces. This project included 6 circuitry/crafting bugs (see Figure 4: three bad connections, two mismatches between pin numbers declared in the code and pins used in the circuit, and one wrong connection) and 20 syntactic coding bugs (see Figure 5: missing delimiters, 4 spaces in variable names, and 6 errors in variable declaration and usage).

When making the DebugIt, Emma, who had little sewing experience, took on the challenge of sewing one circuit across the two different heart-shaped pieces of felt, with lights on one sheet and most of the sewn circuitry on the other sheet (see Figure 3). As she struggled with sewing the microcontroller on one sheet of felt and connecting it to the LED lights on the other sheet, Emma reflected, "Probably I should do more craft," recognizing her need for more sewing expertise (Lesson 3, Video). Yet even knowing this, Emma proposed arranging the electronic components in this spatially complex way and, mischievously, embedding a short circuit between the two layers of felt. Like Gabriel, in the previous case study, Emma took on a challenge that pushed her learning. The group took advantage of the spatial placement of the components and deviously added bugs inside the heart (between the two layers of felt) by placing one LED upside down with the light shining in toward the fabric. The group took on the challenge of making something difficult as an opportunity to improve their skills and design novel bugs.

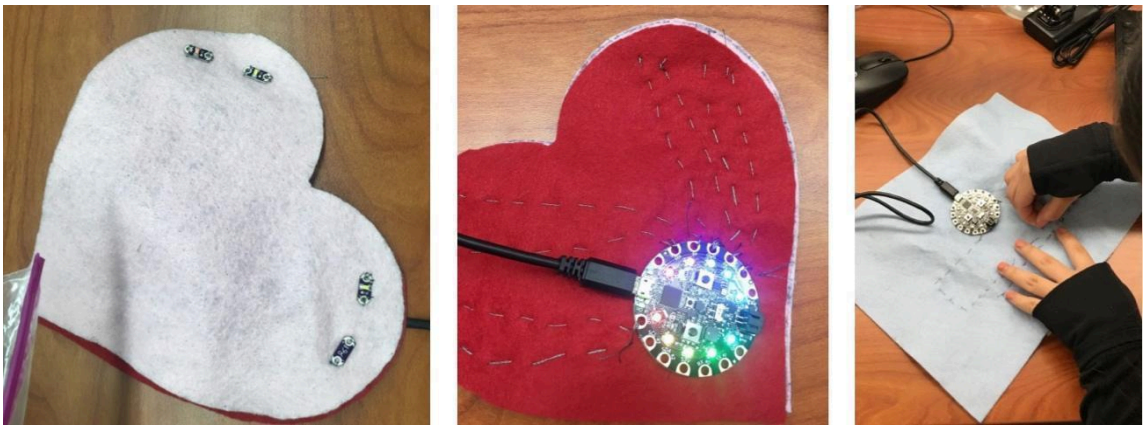

*Figure 4.* Lily, Emma, and Lucas' DebugIt: (1) Left: Front of e-textile: a white heart with four LEDs; one LED is flipped with the light directed towards the felt; (2) Middle: Back of e-textile: the microcontroller is sewn over a second layer of felt; threads are loose; and LEDs are attached to pins 9, 12, 6, and 10; (2) Right: Emma resewing the DebugIt created by her peers.

When choosing what bugs to design in the code, Lucas also chose a challenge that deepened his learning: adding bugs to the conditional statements that evaluated the state of two buttons. This was not an easy task as he was not very familiar with buttons, yet this created an opportunity to make intentional (such as writing "low" instead of "LOW") and unintentional bugs (Lesson 2, Video). While working on this part of the code, he openly shared his coding mistakes, saying, "Oh! I messed up! Wrong in the coding!" as he unintentionally made a semantic mistake by writing the same conditional twice (if(butt1Val == HIGH && butt2Val == HIGH)) when he wanted one to be high for both buttons and the other to be high only for the value of button 1 (Lesson 3, Video).

The group approached learning as constant improvement, as evidenced when, after Lucas finished writing and editing the code, Emma responded, "Let me see what other things we can do," and proposed several ways they could improve the project by adding more bugs, such as assigning the wrong microcontroller pins to program variables (Lesson 3, Video). In a complementary episode, after Emma tested whether the stitches and knots were strong enough by turning the project upside down, Lucas suggested they check the wires to make sure they were not touching, looking for opportunities for improvement (Lesson 3, Video). Lucas and Emma supported the group in finding ways to improve their project, approaching learning as constant improvement.

```
Light 1 = 3;           ▶ Missing data types, variable names should not have spaces, variable
Light 2 = 0;             names do not match those used in the program, wrong pin
Light 3 = 10;            assignments (in the circuit LEDs are connected to pins 6, 9, 10, 12)
Light 4 = 12;
Int button = 4;        ▶ Data type "int" should be lowercase
int button19 = 19;

void setup(){
    pinMode(cap, OUTPUT);
    pinMode(Iman, OUTPUT)      ▶ Missing semicolon
    pinMode(thor, OUTPUT);
    pinMode(wolf, OUTPUT)      ▶ Missing semicolon
    pinMode(button4, INPUT)    ▶ Declared variable is called "button", missing semicolon
    pinMode(button19, INPUT);
}

void loop(){
    int butt1Val = digitalRead(button4);   ▶ Declared variable is called "button"
    int butt2Val = digitalRead(button19);
    if(butt1Val == HIGH && butt2Val == HIGH){ //both buttons on
        side2side();
    }
    else if(butt1Val == HIGH && butt2Val == LOW){ //button 1 on, button 2 on
        flash()    ▶ Missing semicolon                ▲ Mismatch between conditional and comment
    }
    else if(butt1Val == low && butt2Val == HIGH){ //button 1 off, button 2 on
        side2side();
    }                    ▼ Constant LOW should be uppercase
    else{ //both buttons off
        flash();
    }
}

void flash(){
    digitalWrite(cap,HIGH);
    digitalWrite(Iman,HIGH);
    digitalWrite(thor,HIGH);
    digitalWrite(wolf,HIGH);
    delay(100);
    digitalWrite(cap,LOW);
    digitalWrite(Iman,LOW);
    digitalWrite(thor,LOW);
    digitalWrite(wolf,LOW);
    delay(100);
}

void side2side(){
    digitalWrite(cap, HIGH);
    digitalWrite(thor, HIGH);
    delay(300);
    digitalWrite(cap, LOW);
    digitalWrite(thor, LOW);
    digitalWrite(Iman, HIGH);
    digitalWrite(wolf, HIGH)   ▶ Missing semicolon
    delay(300);
    digitalWrite(Iman, LOW);
    digitalWrite(wolf, LOW);
}
```

Figure 5. Code for Lily, Emma, and Lucas' DebugIt annotated with bugs.

While sewing, Emma became frustrated yet overcame plenty of challenges with support from her team. Finding several unintentional problems in her circuit, she reflected and suggested a possible solution: "I always forget to tie the knots; it's so annoying, I better tie them as I go along." Lucas praised her proposed strategy and the process of getting there by saying, "Smart thinking!" (Lesson 3, Video). In a different instance, praising and encouraging effort demonstrated what Love (2019) calls cultural legacy, that is, the traditions and customs of culture, in self-beliefs. Emma had to add a short-circuit bug in which two threads intersected. Her teammates, Lucas and Lily, were worried they would not finish on time. "What time is it? Do you think I'll finish?" Emma asked. In response, Lily and Lucas started singing Dolores Huerta's civil rights chant "¡Sí se puede!"[2] (Yes, we can) to encourage and praise Emma's effort (Lesson 4, Video). Although the team could not finish the circuit that day as class ended, Lily encouraged her peers: "We'll have to be strong to finish" (Lesson 4, Video). They did indeed finish strongly after more work the next day, and their buggy project befuddled the receiving group for a while.

When Lily, Emma, and Lucas debugged a project that their peers had created, we found growth mindset practices just like in all the other groups. For instance, they faced unexpected challenges as they decided to resew the circuit in the project they were debugging—a "sick cloud throwing up a colorful rainbow" designed by Evelyn and Nicolás (for more details on this project, see Fields et al., 2021). The group debated cutting a particularly gnarly-looking thread and resewing the connections (Lesson 6, Fieldnotes). "You could say the biggest challenge was figuring out if we were supposed to cut or leave them there," Lily reflected (Lesson 7, Video). The group was aware that resewing a whole project in a couple of hours was challenging. Yet, by mistakenly taking a feature of the project as a bug, they decided to embark on this challenging task. Emma tried to sew as fast as possible, regularly asking Lily for feedback without being afraid of being wrong or making mistakes. Lily inspected the artifact to make sure the knots were tight and encouraged Emma by telling her that she could fix it. Once Emma finished, Lucas suggested they test the project to make sure the wires did not touch, approaching learning as constant improvement. When testing, they realized the project did not work. They were frustrated and stumped. Yet they persisted and grabbed copper tape to debug the project. "Let's see if it works," Lily said as they tried something new. They covered the stitches on the back of the project with copper tape to make sure the circuit was connected. The copper tape did not solve the issues. As they continued to persist, Lucas and Lily started inspecting the connections to the pins of the microcontroller while Emma inspected the code. In the end, they did not fully debug the "sick cloud" project in the allotted time, something many groups in the class experienced (indeed, it was not a goal for each group to fully debug the buggy projects their peers created). Yet again, this was seen as part of developing comfort with failure, and Lily, Emma, and Lucas continued to learn as they met with the designers and found out some of the bugs they did not diagnose.

---

[2] Dolores Huerta is a Latina civil rights activist who led (with César Chávez) the 1960s farmer and immigrant rights movement in the Southwestern United States. She came up with the chant "¡Sí se puede!" which became a call for action and persistence during rallies (Sowards, 2019).

**Discussion**

In this paper, we analyzed the growth mindset practices that high school student groups exhibited during Debugging by Design (DbD), an intervention created to empower learners to become more familiar, comfortable, and capable with debugging by designing bugs for their peers to solve. Our analysis of these four groups of students provided some insights into how learners developed their growth mindsets and became more comfortable with debugging by attending to the motivational, collaborative, and emotional factors of debugging holistically. Students engaged in practices such as choosing challenges that led to more learning, praising each other's efforts, approaching learning as constant improvement, persisting after setbacks, and developing comfort with failure. This last practice adds to and complements the practices mentioned in Campbell et al.'s (2020) framework of practices for growth mindset learning activities, specifically highlighting mistakes and failures as "an intrinsic part of the learning process" (Papert, 1980, p. 153). While other studies have focused on the conceptual skills and practices students used when debugging physical computing projects or the types of bugs they designed and solved (Lui, Fields, & Kafai, in press), this study focused on the growth mindset practices observed when students were in charge of making or designing bugs. This paper contributes to current research on growth mindset and holistic approaches to debugging by characterizing an inventory of growth mindset practices, highlighting the importance of the design of learning environments in promoting growth mindsets, and providing evidence that collaboration plays an important role in growth mindset practices.

     Our analysis contributes to a more holistic understanding of students' social, emotional, and motivational approaches to debugging through the characterization of growth mindset practices. While it is worth highlighting that most of the growth mindset practices occurred while students were designing bugs, it is necessary to further explore how this compares to other design-based learning activities to be able to draw conclusions about the effect of the intervention. Here, our inventory of practices is a first step that could be helpful in conducting studies that compare mindset practices across interventions and further explore the role that intentionally designing bugs may have on students' mindsets. It could also be helpful for teachers to identify growth mindsets as they develop in their classroom. Future research could expand on where growth mindset practices emerge in other contexts or across a larger set of students. While we attempted to study a diverse range of students (based on the teacher's assessment of which groups had been more and less successful in a previous project), our exploratory study was inevitably constrained by a small sample size (10 students), an eight-class period focus, and small group interactions. Other studies could build on our emergent list of growth mindset practices while also considering other factors, such as how students' prior programming experience and their initial perceptions of computing might also impact their mindsets in *practice* and their development of debugging skills. In the following sections, we consider the designed context for the intervention (Debugging by Design), what we learned about developing students' growth mindset practices, and more directions for further research.

*Debugging and Growth Mindsets in Computing Education*

DbD may provide an opportunity for learners to develop growth mindsets in computing, aligned with Yaeger et al.'s (2018) recognition of learners' needs with an emphasis on giving students a greater sense of control over bugs and debugging. The design of DbD contrasts with previous growth mindset interventions in computing that centered on delivering information about mindsets to students and reflective writing or "saying is believing" exercises (e.g., Simon et al., 2008). With DbD, we aimed to empower students in debugging, with all its cognitive and socio-emotional challenges, through the intentional design of bugs.

DbD built on a longstanding tradition of constructionist activities that put learners in control of their own learning by designing software or game applications for others (Harel & Papert, 1990; Kafai, 1995), extending this concept by shifting the focus from designing functional artifacts (the focus of most instruction) to designing non-functional, or buggy, projects. The learning environment puts students in control of naming and creating mistakes by making them *designers* rather than solely *solvers* of bugs (see Fields et al., 2021). Our analysis provides a compelling example of an "examination of the mindsets conveyed by or embodied" in the learning environment (Dweck & Yeager, 2019, p. 490).

Our analysis revealed that most instances of growth mindset practices occurred primarily when learners designed buggy projects for their peers to solve, rather than later when students solved the exchanged buggy projects. When designing bugs for their peers, learners sought challenges, made and solved unintentional bugs, encountered failure, and persisted. Designing bugs built on adolescents' enhanced desire to be "treated as though they are competent, have agency, and autonomy" (Yeager et al., 2018, p. 104), putting them in control of bugs in interest-driven projects that can challenge their peers. In addition to our earlier research on the benefits of the entire DbD activity, which focused more on students' self-report of gains weeks after the activity was completed (Fields et al., 2021), this paper shows the specific opportunity for practicing growth mindsets in the moment, especially during the *design* phase of DbD.

This has implications for where and how educators might support high school students' growth mindsets in action (i.e., in practice). As previously discussed, most research studies on growth mindsets in computing take top-down, didactic approaches that may result in false growth mindsets—that is, simply telling students about growth mindsets without providing them with strategies and opportunities for practice (Dweck & Yeager, 2019). While explicit instruction and reflection on growth mindset can certainly help students (Burnette et al., 2020; Kwak et al., 2022), we suggest that this design-focused, practice-first approach may be particularly well suited to computing education. Learning environments such as DbD may help promote growth mindset in practice in observable and concrete ways, giving students the opportunity to encounter or design for failure in order to develop resilience and take ownership of their own learning. While our research does not make a causal claim, this approach of putting learners in charge of designing mistakes for their peers to solve and solving bugs in their peers' projects requires further investigation. This may involve comparing the practices of students in DbD with those of students creating open-ended projects. DbD could also be applied to other areas of computing education that involve iterating, testing, and solving problems.

*Towards a Holistic Approach to Debugging*

We sought a holistic approach in studying growth mindset practices while students designed and solved bugs, emphasizing the motivational, emotional, and social aspects of debugging. The case studies demonstrated how debugging involved cognition and emotion (DeLiema et al., 2020) and not just basic skills for identifying and solving isolated problems. When encountering cognitive challenges, student motivation and emotion influenced how they addressed bugs, and at the same time, bugs themselves affected motivation and emotion. Further, while both debugging and mindset practices are often discussed as individual processes, collaboration also plays a key role. Accepting debugging as a holistic process requires us to examine the role of social interaction and culture in addition to cognition and emotion.

Peer collaborations played a critical role as learners engaged with growth mindset practices during DbD. As we saw in our analysis, 78% of the instances of growth mindset practices identified occurred in peer-to-peer interactions. Furthermore, peer feedback and support were critical in promoting persistence, reshaping setbacks as opportunities for learning, and providing a space for students to support one another when faced with failure and frustration. This aligns with other studies on debugging e-textiles that show the importance of providing opportunities for peer collaboration (Jayathirtha et al., 2020; Shaw et al., 2019; Lui et al., in press). It also aligns with earlier research on DbD that showed how learners felt more comfortable asking peers for help after doing the DbD activity (Fields & Kafai, 2020).

More broadly, one unanticipated finding is that debugging and growth mindset practices are embedded within cultural discourses present in learners' communities. In our analysis, we saw how student cultural legacy manifested in learners' beliefs about their ability to debug when Lily and Lucas started chanting Dolores Huerta's "¡Sí se puede!" to encourage and celebrate their peers' efforts. The "¡Sí se puede!" chant, which emerged in the 1970s during the civil rights movement for farm and immigrant worker rights in the Southwestern United States, is an expression of encouraging effort and persistence with great cultural and historical significance in Latinx communities (Sowards, 2019). Indeed, as Love (2019) writes, perseverance for justice, civil rights, and survival among minoritized communities are manifestations of incremental theories of intelligence (such as growth mindset or grit) that are usually ignored in the classroom but that ought to be recognized, protected, and nurtured. This aligns with other research that demonstrates how students making e-textiles can leverage culturally relevant practices (Kafai et al., 2014) and develop their sociopolitical identities (Shaw et al., 2021). Further research could address how debugging interventions can build on the cultural legacies of students and take advantage of the growth mindset practices that are already part of their communities. A truly holistic approach to debugging must include not only cognition, emotion, and motivation but also culture and social interaction.

Finally, the collaborative, holistic work in this intervention happened in a socially situated classroom context. The design of the DbD unit was part of that context, but so was the highly experienced teacher who promoted a classroom culture throughout the school year that encouraged student collaborations, legitimized student expertise, and modeled comfort with failure (for more information on teaching, see Fields et al., 2019; Fields et al., 2021). While our analysis in this paper focused on case study student groups, the teacher played a crucial role by intentionally grouping students with different abilities, providing just-in-time feedback, and encouraging student groups to design more challenging bugs and explore computing concepts beyond what was covered in the class. Future work could explore holistic supports for growth mindsets at the classroom level.

*Future Directions*

We reported on emergent growth mindset practices exhibited by high school students in the first classroom-based implementation of DbD. Given the lack of research on K–12 students' growth mindsets in computing, much more work could explore growth mindsets in other contexts of debugging and more general computing learning. Such research should focus on the social, motivational, and emotional aspects of creating and solving bugs, as well as how teachers can foster supportive collaborative environments in which students are not afraid of bugs or seeking help and support from their peers. In future studies we plan to research how different teachers promote growth mindset practices when implementing or integrating the activity in their computing classrooms. This will allow us to examine the range of teaching practices that support growth mindset in the context of DbD. The case study analysis already provided some evidence of how teachers played a role in students' engagement in growth mindset practices through feedback and advice at key points of the design process. Comparing student growth mindset practices across classrooms and teachers implementing the activity may help identify what role DbD has, if any, in promoting growth mindset.

    Additionally, it is necessary to analyze the relationship between growth mindset practices observed in the classroom and the perceived growth mindset of students and teachers, as assessed through self-reported instruments. We recently developed and validated a pre- and post-survey instrument to measure student self-beliefs in computing and debugging in general, as well as in e-textiles, based on the implementation discussed in this paper (Morales-Navarro et al., 2023a). We used this instrument in a quasi-experimental study, with some classrooms doing DbD and others designing an additional e-textiles project. Results from this other study suggest that DbD project completion was uniquely correlated with increased programming growth mindset (Morales-Navarro et al., 2023b). Learners can benefit not only from designing functional applications, but also from designing buggy applications—both of which are and should be an intractable part of any learning production.

    Beyond DbD and e-textiles, our growth mindset practices inventory may be useful for observing growth mindset in action in other computing education settings. Similarly, future research could investigate the relationship between growth mindset practices and debugging practices and strategies that students engage with when solving (or creating) bugs in open-ended projects.

**Conclusions**

This paper illustrated different growth mindset practices that high school students engaged with during an intervention designed for students to create buggy projects for their peers. Analysis of students' peer interactions and case studies showcased how high school student teams designed bugs for others to solve and in the process chose challenges that led to more learning, persisted after setbacks, gave and valued praise for effort, approached learning as constant improvement, and developed comfort with failure. These practices can be helpful to further study growth mindset in action in other computing learning environments. Computing educators and researchers must strive to develop interventions that engage students in developing growth mindset practices rather than simply telling them about them. Finally, the idea of making students designers of bugs, or mistakes, is not limited to e-textiles. Debugging by Design could be used as a pedagogical approach in other areas of physical computing and programming education in general. We look forward to what future applications of DbD can show about student learning and engagement with growth mindset practices.

**Acknowledgments**

This work was supported by a grant from the National Science Foundation to Yasmin Kafai (#1742140) and Michael Eisenberg (#1742081). Any opinions, findings, and conclusions or recommendations expressed in this paper are those of the authors and do not necessarily reflect the views of NSF, the University of Pennsylvania, the University of Colorado, Boulder, or Utah State University. Special thanks to Lindsay Lindberg and Ammarah Aftab for their support in data collection and analysis.

**Research Ethics and Consent Statement**

We recruited students already enrolled in an introductory computing high school class. A researcher visited the class to invite students to participate, distribute consent and assent forms, and address any questions about the research procedures. Parents received consent forms prior to the study, which included a brief explanation of the research, and youth assented to their participation in writing. Students did not receive any incentives for participating in the study. The IRB board of the University of Pennsylvania approved research protocols and data collection techniques (Protocol: 827747).

**Data Availability Statement**

Due to the nature of this research, participants in this study did not agree for their data to be shared publicly. Supporting data is not available.

**Disclosure Statement**

No potential conflict of interest was reported by the authors.

**Funding details**

This work was supported by the National Science Foundation [1742140].

**Appendix 1.**

Coding Scheme:

| Code and Definition: | Example 1 | Example 2 |
|---|---|---|
| **Choosing challenges that lead to more learning:** Students purposefully choose challenges that lead them to learn new things saying things like "I like a challenge" instead of avoiding challenges that may expose areas of weakness. | Despite her little experience sewing ["probably I should do more craft"], Emma decides to sew the microcontroller and the LEDs on two sheets of felt one after the other. This requires her to turn the heart to figure out how sewing the CP on one piece of felt and the LEDs on the other would work. (Videolog Group 1 Day 3, page 7) | When coming up with bugs, Evelyn says she wants to include music. She explains that she wants to press a button to activate the music. Nicolás says that's hard and that he doesn't know how to code music. Then Nicolás turns to his computer and starts typing. A 10 minutes later and after trying things on the computer, Nicolás shows Evelyn how they could add music (Videolog Group 3 Day 2, page 3). |
| **Persisting after setbacks - with preceding frustration:** Students display frustration and persist, overcoming it with continued engagement with the problem at hand. | Evelyn looks at the heart and points out that "they (the designing group) didn't tie any knots." She seems frustrated. Suddenly she manipulates the heart and exclaims, "I did it." (Videolog Group 3 Day 2, page 3) Despite her frustration she keeps sewing. Evelyn "I could have fixed it a long time ago if I knew the connections were all loose" she reflects. (Videolog Group 3 Day 7, page 5) | Evelyn has been trying to work on an aesthetic and circuit design. She says everything is too hard to draw. Nicolás suggests using squares and circles to start and then adding more basic shapes to represent what they want. (Videolog Group 3 Day 2, page 3) Evelyn follows this new strategy and finishes the drawing. (Videolog Group 3 Day 2, page 4) |
| **Persisting after setbacks - co-occurrence of frustration:** Students encounter a challenge and overcome it by trying again or adopting a new strategy: "I'll have to try harder or work differently." | Evelyn is frustrated with a bug that prevents the lights from working, her partner encourages her "You can fix it! Fix it!," she tries again and manages to get the lights to turn on (Videolog Group 3 Day 7, page 2) | Gabriel cannot manage to get the music to play, he has a function that plays it and is calling it, but it never plays. He tries calling the function in the void setup() to get the music to play (Videolog Group 2 Day 7, page 7) |
| **Giving and valuing praise for effort:** Students praise their own or their peers' efforts, strategies, and processes | Camila asks Georgia: "it looks good, right?" Showing her the note she just cut. "Yeah, it looks real good" Georgia replies. (Videolog Group 2 Day 3, page 9) | Nicolás praises Lily for adding nuisance code [declaring four extra variables that were not used] just to confuse the debugging team, saying "good job, you made it really hard [for the other team]" (Videolog |
| | | Group 3 Day 7, page 7) |
| **Approaching learning as constant improvement:** Students see the goal of learning as improving performance. | Lucas finishes writing the code and says "Success" indicating he implemented the code changes Emma suggested. Emma responds, "Let me see what other things we can do" and looks for ways they can improve their project. (Videolog Group 1 Day 3, page 3) | When deciding on how to organize the negative and positive threads in their project, Nicolás reflects "we should have done this for the mural project. It would have been nice" as he admires how their current strategy for the design could have improved their mural project. (Videolog Group 3 Day 4 page 2) |
| **Developing comfort with failure - (1) asking questions and requesting feedback without fear of being wrong:** Students ask questions to their peers with no fear of being wrong. | Liam asked Sophia if the constant pin values in Arduino ("LOW" and "HIGH") could be lowercase. (Videolog Group 4 Day 3, page 8) | While sewing a circuit Sophia asked Liam "Can positive and positive cross each other?" (Videolog Group 4 Day 3, page 7) |
| **Developing comfort with failure - (2) sharing failure with their peers:** Students recognize errors or mistakes and share them with their peers. | Georgia realizes she sewed the LED to the wrong pin. "Was I supposed to sew number six" Camila "I don't know" Georgia "I sewed it to number ten. (Videolog Group 2 Day 3, page 9) | Evelyn shared a mistake "Oh my god! I was going to keep going without sewing up the light!" (Videolog Group 3 Day 4, page 2) |
| **Developing comfort with failure - (3) embracing failure and imperfection as part of the process:** Students voice that failure is part of the process and that their projects do not have to be perfect. | When the felt doesn't align Emma explains "it's a little bit off, but it's okay". She offers a solution, saying "we'll just cut it off, it doesn't have to be perfect" (Videolog Group 1 Day 4, page 6) | Georgia "I'm just warning you I'm not the best at sewing, that's alright. As long as it all looks okay at the end it doesn't really matter." (Videolog Group 2 Day 3, page 6) |